\documentclass[draft]{agujournal2019}
\usepackage{url} 
\usepackage{lineno}
\usepackage[finalnew]{trackchanges}
\usepackage{soul}
\usepackage{ragged2e}
\linenumbers
\addeditor{editor one}

\draftfalse

\journalname{Space Weather}

\begin{document}
\nolinenumbers
%
%

\title{Modeling the dynamic variability of sub-relativistic outer radiation belt electron fluxes using machine learning }

%
%




\authors{Donglai Ma\affil{1},Xiangning Chu\affil{2},Jacob Bortnik\affil{1},Seth G.Claudepierre\affil{1,4},
W.Kent Tobiska\affil{3},Alfredo Cruz\affil{3},S.Dave Bouwer\affil{3},J.F.Fennell\affil{4},J.B.Blake\affil{4}}

\affiliation{1}{Department of Atmospheric and Oceanic Sciences, University of California, Los Angeles}
\affiliation{2}{Laboratory for Atmospheric and Space Physics, University of Colorado Boulder, Boulder, CO, USA}

\affiliation{3}{Space Environment Technologies, Pacific Palisades, CA, USA
}
\affiliation{4}{Space Sciences Department, The Aerospace Corporation, El Segundo, CA, USA
}





\correspondingauthor{Donglai Ma}{dma96@atmos.ucla.edu}




\begin{keypoints}
\item A set of neural network models was developed to reconstruct the 50 keV-1 MeV electron fluxes in the outer radiation belt.

\item The models reproduce fluxes with a high overall accuracy ($R^2 \sim$  0.78-0.92) during out-of-sample periods, including long- and short-term dynamics.
 
\item The models reproduce the time-varying MLT dependence exhibited during storms by the lower energy electrons ($\sim <$100 keV).
\end{keypoints}

%
%

%
%


\begin{abstract}
      \justifying
We present a set of neural network models that reproduce the dynamics of 
electron fluxes in the range of 50 keV $\sim$ 1 MeV in the outer radiation
belt. {The Outer Radiation belt Electron Neural net model for Medium energy electrons (ORIENT-M) uses only solar wind conditions and geomagnetic indices as input.} The models are trained on electron flux data from the Magnetic Electron Ion Spectrometer (MagEIS)
instrument onboard Van Allen Probes, and they can reproduce the dynamic variations of 
electron fluxes in different energy channels. The model results show high coefficient of determination ({$R^2 \sim $} 0.78-0.92)
on the test dataset, an out-of-sample 30-day period from February 25 to March 25 in 2017,
when a geomagnetic storm took place, as well as an out-of-sample one year period after March 2018. 
In addition, the models are able to capture electron dynamics such as 
intensifications, decays, dropouts, and the Magnetic Local Time (MLT) dependence of the lower energy ($\sim <$ 100 keV ) electron fluxes
during storms. The models have reliable prediction capability and can be used for a wide
range of space weather applications. The general framework of building our model is not limited to 
radiation belt fluxes and could be used to build machine learning models for a variety of other 
plasma parameters in the Earth's magnetosphere.

\end{abstract}

\section*{Plain Language Summary}
\justifying
The Earth's radiation belts consist of energetic particles trapped by the geomagnetic field. This radiation environment is known to be particularly hazardous to spacecrafts and difficult to predict given the complex dynamics of electrons at different energy states. This paper presents a set of neural-network-based models that use measurements of geomagnetic and solar activities as drivers to reconstruct radiation belt electron fluxes ranging from 50 keV to 1 MeV. The models can determine the flux with high accuracy and capture the electron dynamics with long- and short-term time scales. The models provide reliable prediction capability and can be used for a wide range of space weather applications. The approach through which our models are built is not limited to radiation belt fluxes and can be generalized for a variety of other plasma parameters in the Earth's magnetosphere.

%
%

\section{Introduction}
\justifying
The Earth's energetic particle environment consists of electrons that range in
energy from a few keV to multiple MeV \cite <e.g.>{Baker2017,Thorne2010}, and
were discovered by Geiger counters flown on Explorer 1, launched in January 1958,
which represents the first major discovery of the space age
\cite{VanAllen1958,VanAllen1959}. The dynamics of the different electron energies that
have been studied over several decades, but more recently in great detail facilitated by
high quality data from the Van Allen Probes mission \cite{Mauk2013,LiandHudson2019}. The 
exact response of the radiation belts to solar wind driving is difficult to predict due to
various competing loss and acceleration processes, often giving vastly different responses
to seemingly similar driving conditions \cite{Reeves2003} but occasionally even being
accelerated up to roughly 10 MeV. The generally accepted process of radiation belt flux 
enhancement is believed to proceed as follows: an increase in the strength of the convection
electric field causes an enhanced drift of plasma-sheet electrons into the inner
magnetosphere, where they gradient-drift eastwards, and become unstable
to plasma wave excitation. A particular class of waves, called whistler-mode chorus waves
is excited on the dawn side of the Earth by $\sim$10-100 keV ‘source’ electrons, and these
waves then transfer a portion of their energy to higher energy ‘seed’ electrons ($>$100 keV),
that are further accelerated to relativistic energies ($\sim$MeV) facilitated by Ultra-Low
Frequency (ULF) waves \cite{BortnikandThorne2007,Xinlin2006,Reeves2003,Thorne2013,Turner2013,Xiao2009,LiandHudson2019}.

In addition, the energy-dependent dynamics of radiation belt electrons at multiple energies
can operate on different time scales and have vastly different behaviors \cite<e.g.>{Reeves2016,Ma2016}.
For example, the Van Allen Probes were able to identify an "impenetrable" barrier to relativistic electron
inward motion \cite{Baker2014} at L $\sim$2.8, while lower energy electrons exhibit fast dynamics and are more
likely to fill the slot region and penetrate into the inner zone \cite{Reeves2016}. Besides, more enhancement
events occur at lower energies compared to higher energies while the lifetime of the former is much shorter than
the latter in the outer radiation belt \cite{Seth2020b,Seth2020a,Reeves2016}. Thus, developing useful models to
describe the trapped electron flux across a range of different energy channels is important for understanding and predicting 
the dynamics and distribution of the radiation belts.

The outer radiation belt electron fluxes and their dynamics have been modeled using a variety of approaches including
both data-driven and physics-based models. The empirical AE8 and AE9 models are statistical models which are static, and reconstruct the
distribution of electrons from 40 keV to 7 MeV \cite{ae9,sawyer1976ap}. Other statistical models such as 
CRRESELE \cite{bell1995} covers 500 keV $\sim$ 6.6 MeV, the IGE-2006 \cite{ige2006} and the POLE \cite{POLE} are applicable only for geostationary orbit,
whereas the FLUMIC \cite{wrenn} and the MODE-DIC model \cite{hands2015} model the worst case scenario for 1-day fluence and environment for spacecraft internal charging.
These empirical models depend on L shell and energy but are usually independent of geomagnetic activity, meaning
that they are static, i.e., they reproduce 'statistically averaged' conditions. Several machine learning models have also been developed to study
trapped electron fluxes in the radiation belts. These include \citeA{balikhin2011} who used the Nonlinear Autoregressive Moving Average with Exogenous inputs (NARMAX) approach,
and \citeA{zhang2020} who used Artificial Neural Networks (ANNs) with quantile regression. Both models focus on the fluxes at geosynchoronous (GEO) orbit. \citeA{Smirnov} adopted a gradient boosting decision tree (GBDT)
method on GPS data to model the fluxes of electrons with energies in the range 120-600 keV in the MEO region which only captures one altitude.
\citeA{PIRES} used a combination of data collected at low earth orbit and LANL-01A at GEO orbit and POES electron fluxes as the input of 
the model to predict MeV energy flux and \citeA{seth2020} also included POES data from LEO as the driver of their neural network model, which operates on a daily cadence.

Using physics-based models is another important way to replicate and understand the dynamic behavior of the radiation belts. One common method is through the integration of 
the Fokker-Planck (FP) equation \cite{Lyons1973EquilibriumSO,Glauert2014,ma2015,Reeves2012,tu2013}. However, the FP model can only represent the effects of diffusive scattering on the particles,
it cannot represent nonlinear processes, nor include any unknown (or unquantified physical process which might shift the balance between acceleration and loss processes).
And such physical-based modeling requires artificial or observed initial and boundary conditions to drive the simulation
which also need to be specified during the course of the simulation.

Here, we present a novel strategy based on the Artificial Neural Network (ANN) model to overcome some of these shortcomings. This approach is versatile and can be applied to a
variety of physical quantities sampled at sparse locations and times, and has already been utilized in space weather modeling the plasma density, chorus and hiss waves, and in conjunction with physics-based radiation belt modeling
\cite{Bortnik2016,chu2017den,chu2017refill,BORTNIK2018279,chu2021}. We describe and demonstrate the development of several ANN models of outer radiation belt electron fluxes
covering energy channels from $\sim$50 keV to $\sim$1 MeV based on Van Allen Probes data. Since solar
wind driving and geomagnetic activity alone determine the fluctuation of the outer radiation belt, the neural network model accepts these parameters as inputs without
relying on any boundary conditions from other satellite data, making this model a tool with great potential in predicting and understanding radiation belt dynamics.
We show that our electron flux model captures the dominant features of the variability at different time scales, as well as the energy-dependent dynamics and the MLT dependence of the 
low energy channel.

\section{Data Description}

\subsection{MagEIS Electron Flux data}
The primary data used throughout this work are obtained from the Magnetic Electron Ion Spectrometer (MagEIS) instrument suite
\cite{Blake2013} onboard NASA's Van Allen Probes \cite{Mauk2013}. The identically instrumented twin-spacecraft Van Allen Probes 
mission has a highly elliptical low-inclination orbit with an apogee of $\sim 6  R_E$ and a perigee of $\sim$ 600 km. The MagEIS instruments measure electron fluxes throughout this
orbit over a wide energy range ($\sim$30-4 MeV) with four electron spectrometers (LOW:$\sim$30-200keV, M75, and M35:$\sim$200 keV to 1 MeV, HIGH:~1-4MeV). The current study illustrates the capabilities of the 
machine learning model using four selected energy channels from the MagEIS instrument (54 keV, 235 keV, 597 keV, and 909 keV shown in Figure 1). Extensive background correction of MagEIS spin-averaged fluxes
was performed to remove background contamination due to inner belt protons and high energy electrons that produce bremsstrahlung X rays \cite{seth2015,seth2019}. However, there are periods when the background correction cannot be performed,
and as a result, background corrected MagEIS data at energies $< \sim$220 keV is only available at L $>$ 4 on RBSP-B. For lower energy electron fluxes, the drift period is relatively
long. Thus the data available on both probes are preferentially used in our study in order to capture the magnetic local time (MLT) dependence. For the above reasons, we use uncorrected data on 54 keV 
channel and background corrected data on other three channels.
\begin{figure}
      \noindent\includegraphics[width=\textwidth]{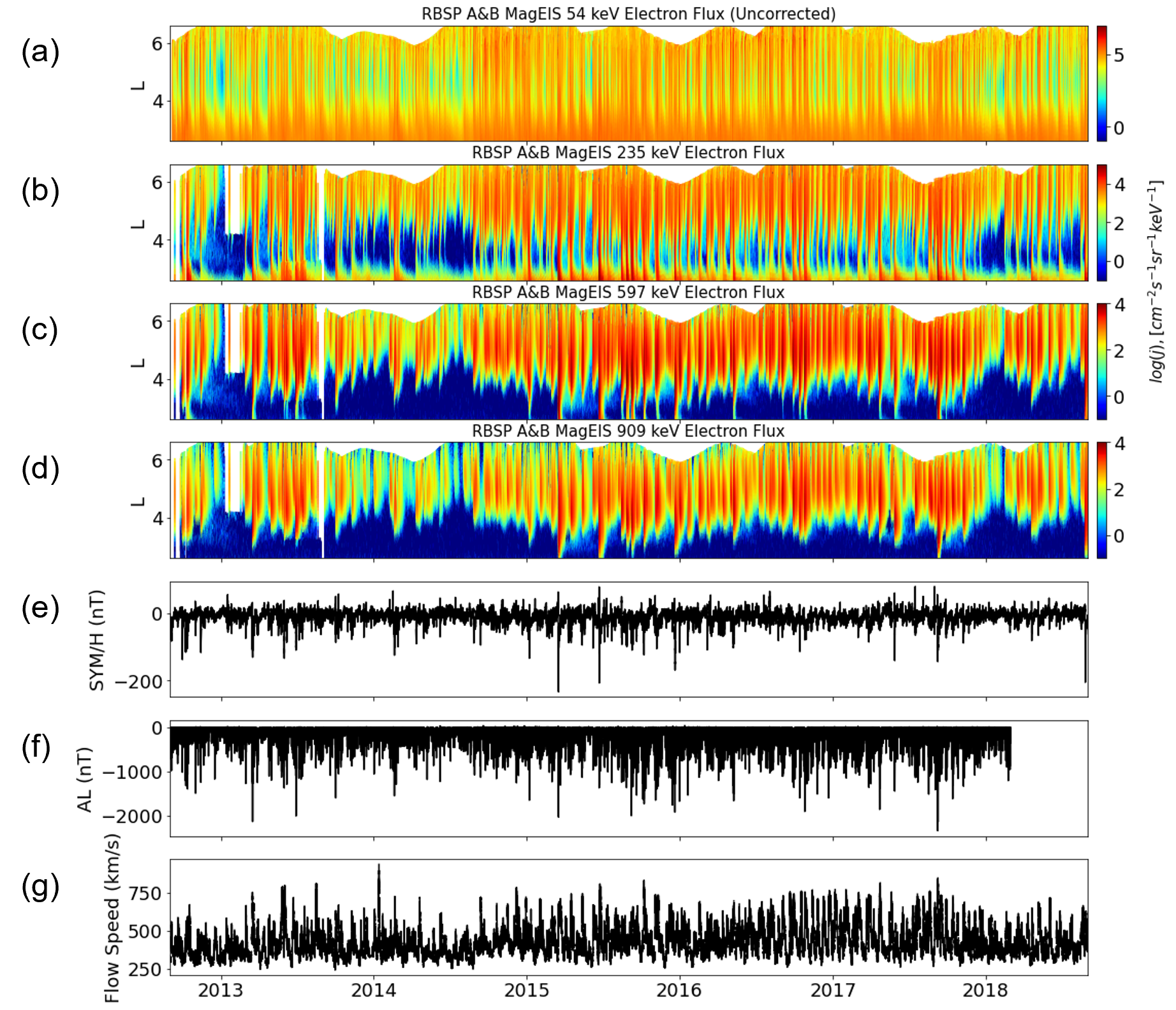}
      \caption{ (a-d) Spin-Averaged fluxes of electrons with energies of 54, 235, 597, 909 keV from MagEIS
      on RBSP A and B for the indicated time interval. The MagEIS fluxes reflect uncorrected
      data for channel 54 keV and corrected data for the remaining channels. (e-g) 5-minute resolution SYM-H
      and AL indices and solar wind speed from the OMNI dataset.}
      \label{Figure1}
      \end{figure}
The temporal resolution of the data is reduced to 5 min averages, and the L shell is restricted to L $>$ 2.6 to focus on the dynamic variability in the outer radiation belt, which results in a data set containing $\sim$1 million data points in each
energy channel.
\subsection{OMNI dataset}
The variation of the Earth's outer radiation belt is known to be driven by the solar wind (e.g., solar wind pressure leading to 
magnetopause shadowing, inward convection of plasma from the tail, and creation of ultra-low frequency waves facilitating radial diffusion) and the resultant
geomagnetic activity (e.g., whistler-mode chorus and hiss waves induced by the plasma injections from the magnetotail). Therefore, the neural network model uses both solar wind parameters as well
as geomagnetic indices acquired from the OMNI data set (\url{https://omniweb.gsfc.nasa.gov/}) as input parameters (5-minute resolution). It is important to use both because occasionally structures observed in the solar wind 
can miss the Earth resulting in minimal geomagnetic activity, or vice versa, so using both solar wind parameters and geomagnetic indices reflects both fundamental drivers and their geoeffectiveness \cite{walsh2019}.
The OMNI AL and AE indices end after March 1 2018, so the electron flux data used in the model training process is from October 2012 to February 2018. The candidate input parameters include AE, AL, AU, ASY-D, ASY-H, Bz-GSM, E, flow speed, solar wind pressure, proton density, SYM-H, and SYM-D.

\section{Methodology}
\subsection{Model description}
We use a simple {fully-connected Multi-Layer Perceptron (MLP)} model to reconstruct the energetic electron fluxes from a number of different channels throughout the 
outer radiation belt. The approach represents an extension of the technique used in a number of previous studies, which successfully modeled plasma density, waves, and electron fluxes \cite{Bortnik2016,BORTNIK2018279,chu2017den,chu2017refill}. The input parameters used for each neuron in
layer $l$ are the products of the output of the preceding layer's $(l-1)$ nodes with their associated weights

\begin{equation}
z_{j}^{l}=f\left(\sum_{i=0}^{N-1} z_{i}^{l-1} w_{i j}+b_{i}\right)
\end{equation}
where $i$ and $j$ denote the neuron number in the preceding $(l - 1)$ and current $(l)$ layers, respectively, and $w_{ij}$ and $b_i$ are
the weights and biases in the hidden layer. The output of each neuron in the hidden layers is calculated using a RELU activation function

\begin{equation}
f\left(z^{l}\right)=\max \left(0, z^{l}\right)
\end{equation}
which is one of the most widely used activation functions in deep learning \cite{lecun2015deep,nair2010rectified}.
After each hidden layer, we place a batch normalization layer to prevent the vanishing and exploding gradient problem \cite{ioffe2015batch}.

The outputs of the preceding $(l-1)$ layer's neurons are all used as the inputs of the neurons in the next layer $(l)$,
thus creating a fully-connected feedforward neural network. Since the corrected MagEIS electron fluxes has many zero values, the model takes the 
function $log_{10} (flux + 1)$ as its output, where the 1 is added to avoid taking the log of zero, but does not affect the flux value otherwise since flux values typically take on large values $\sim 10^4 - 10^6$ as shown in figure 1. The model is trained using the Nesterov-accelerated Adaptive Moment Estimation (Nadam) optimizer with default settings in Tensorflow (\url{https://www.tensorflow.org/api_docs/python/tf/keras/optimizers/Nadam}) to minimize the mean squared error (MSE) of the output.
The dropout and early stopping methods are used in the same way in previous studies \cite{Bortnik2016,BORTNIK2018279,chu2017den,chu2017refill,chu2021}. 

{The whole dataset of electron flux measurements from RBSP A and B (October 2012 - April 2019) was split into three distinct subsets: Part 1: One-year interval between March, 2018 and April, 2019 when the AL and AE indices from the OMNI dataset were not available; Part 2: One-month interval between February 25 and March 25, 2017, when a geomagnetic storm occurred; Part 3: Remaining data. We used Part 3 to train, validate and test the model. Part 1 and Part 2 were only used to evaluate the model's true out-of-sample performance.
The Part 3 was further split into 2-day segments. The 2-day segment length was chosen so that it would be much larger than the time resolution of the electron fluxes and input parameters, to ensure that the data samples that are close in time will not be correlated between testing and training sets, and it would be much smaller than the whole period of five years to ensure a large number of time chunks and that target flux data in the testing and training sets were distributed similarly. \change{It is worthwhile noting that in the same 2-day interval, which includes both RBSP A and B observations, merged together will be divided into the same data segment. Therefore, we make sure that the observations of RBSP A and B from the same time intervals will be retained as either training or validation data, not both, to avoid data leakage.}{It is worth noting that in each 2-day interval, both RBSP A and B observations are merged together from the outset and then divided into separate data segments. Therefore, we ensure that the observations of RBSP A and B from the same time intervals will be retained together in either the training or test data set, will never appear in both, in order to avoid data leakage.}

The training process of the final model includes two stages:
Stage 1: Input parameter selection. In this stage, we use a 5-fold cross-validation method, a subsample of 20\% segments are retained as the validation data for testing the model, and the remaining 4 subsamples are used as training data. The cross-validation process is then repeated 5 times, with each of the 5 subsamples used exactly once as the validation data to get the most accurate quantification of the errors. The detailed process of input parameters selection is described in Section 3.2; 
Stage 2: Hyperparameter optimization for a best model. After we determine the input parameters, we optimize the hyperparameters to obtain a best final model, with 70\% of segments used for the training set, 15\% for the validation set and 15\% is set side as the test set to evaluate the model performance.} The training process uses the training dataset to update the weights and biases, and its performance is assessed on the validation dataset.
When the MSE of the validation set stops improving for several steps in a row (8 steps in this study), the training process ends to avoid
overfitting and ensure maximum generalizability to unseen data. 
\subsection{Input parameters selection and hyperparameters tuning}
\begin{figure}
      \noindent\includegraphics[width=\textwidth]{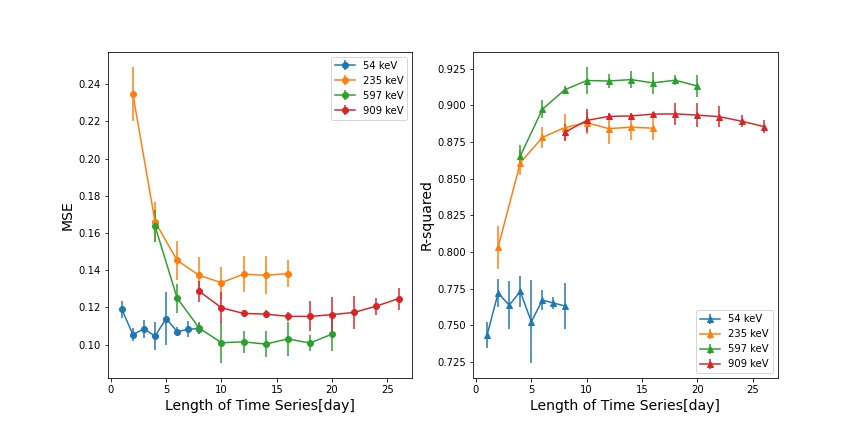}
      \caption{ Model performance on the test {dataset} using different lengths of time series as input parameters for each channel.
      The error bars show the median and standard deviation of MSE (left panel) and coefficient of determination R-squared.}
      \label{Figure2}
      \end{figure}
Our ANN models can be described as $F_i(\mathbf{X}(t);\mathbf{Q}_{1i},\mathbf{Q}_{2i},...,\mathbf{Q}_{ki})$, where $i$ denotes the energy channel being modeled, $k$ denotes the selected 
different input indices, and $\mathbf{X}(t)$ represents the position of the spacecraft at the time of observation, using the vector $(L$, $\sin(MLT)$,$\cos(MLT)$,$MLAT)$. The reason for using trigonometric 
functions for the azimuthal coordinate is to eliminate the discontinuity at $MLT = 0$. The $\mathbf{Q}_{ki}(t,t_{0i},t_{hi})$ is time series of 
time-averaged candidate solar wind parameters or geomagnetic indices with resolution $t_{0i}$ and time history of length $t_{hi}$. The set of time resolution values $t_0$ for the 909 keV channel was based on our previous 1.8 MeV model \cite{chu2021} and decreasing with energy based on the different characteristic dynamic time scales. The initial $t_h$ is based on the electron lifetime observed by \citeA{Seth2020a} for different channels and is further tuned in the following step.

The feature selection technique utilized here is essentially focused on adding the most informative predictors to the model sequentially \cite{kuhn2013applied}
 and evaluating its performance. First, We select a 3-hidden-layer neural network model with 500, 100, and 10 neurons respectively, and with 0.2 drop rate for each layer as a base model. The model size selection is based on our total sample size and previous study experience. We loop through the candidate $\mathbf{Q}$ in Section 2.2 as input and evaluate the test performance using group 5-fold cross-validation. After we iterate through all the input parameters $\mathbf{Q}$, we select the time series with best performance (e.g. AL), and then combine that $\mathbf{Q}_{best}$ with another $\mathbf{Q}$ to repeat the iteration process. Finally, we stop the iteration when $k = 4$ for all the channels due to consistency and the test results that improvements of $R^2$ score value at $k = 5$ are small (less than 0.01). The specified parameters $\mathbf{Q}_k$ for each channel are shown in Table 1 in decreasing order of importance. The fact that final parameters include solar wind speed ($V_{sw}$), pressure ($P_{sw}$), SYM-H and auroral electrojet (AE or AL) index is consistent with our physical understanding of the processes involved and a large number of previous studies \cite<e.g.>{Baker1979,Baker2014,simon,simonnew}. 

 We then studied the model performance at various time periods to discover the ideal length $t_h$ for the time series of input parameters. Figure 2 (left) plots the median and standard deviation of the test performance based on the group 5-fold cross-validation, while the right panel shows
 those of the coefficient of determination, denoted as R-squared. The best $t_h$ values determined for each channel are shown in Table 1. They are considerably longer than the estimated electron lifetime in the outer radiation belt for the corresponding energy \cite{Seth2020b,Seth2020a},
 which suggests that dynamics with time scales smaller than simple decay would be captured in the input time series. After we determined the input parameters, we retrain the models as described in Section 3.1 {Stage 2.} { The hyperparameters of the ORIENT-M models, including the number of hidden layers, the number of neurons in each hidden layer and the dropout rates, are optimized using a Tree-structured Parzen estimator algorithm (Bergstra et al., 2011, 2013) implemented in Optuna (Akiba et al., 2019). The final models' structure and Optuna's hyperparameter search space are shown in Supplementary. }We are aware of the limitations that the tuning process of the hyperparameters should be performed at once, including the time-lags of the input parameters, the hyperparameters of the neural networks and etc. However, the number of hyperparameters are so large that the search space of hyperparameters is huge and will require a very large amount of computation power. Therefore, the workflow has been adapted as described above to save computation time. \add{A complementary method which could help this complicated parameter selection problem is using information theory to investigate the information transfer from one parameter to another and the response lag time }\cite<e.g.>{simon,simonnew}.

\begin{table}[]
      \centering
      \caption{Selected lengths of time history $t_h$, time resolution $t_0$ and parameters $\mathbf{Q}_k$ at different channels}
      \begin{tabular}{cccc}
      \hline
      Channel & $t_0$   & Best $t_h$ & $\mathbf{Q}_k$               \\ \hline
      54 keV  & 30 min  & 4 d        & AE, SYM-H, Psw, Vsw \\ \hline
      235 keV & 60 min  & 10 d       & AL, Psw, SYM-H, Vsw \\ \hline
      597 keV & 90 min  & 14 d       & AL, SYM-H, Psw, Vsw \\ \hline
      909 keV & 120 min & 18 d       & AL, SYM-H, Vsw, Psw \\ \hline
      \end{tabular}
      \end{table}

\section{Model results} 
\begin{figure}
      \noindent\includegraphics[width=\textwidth]{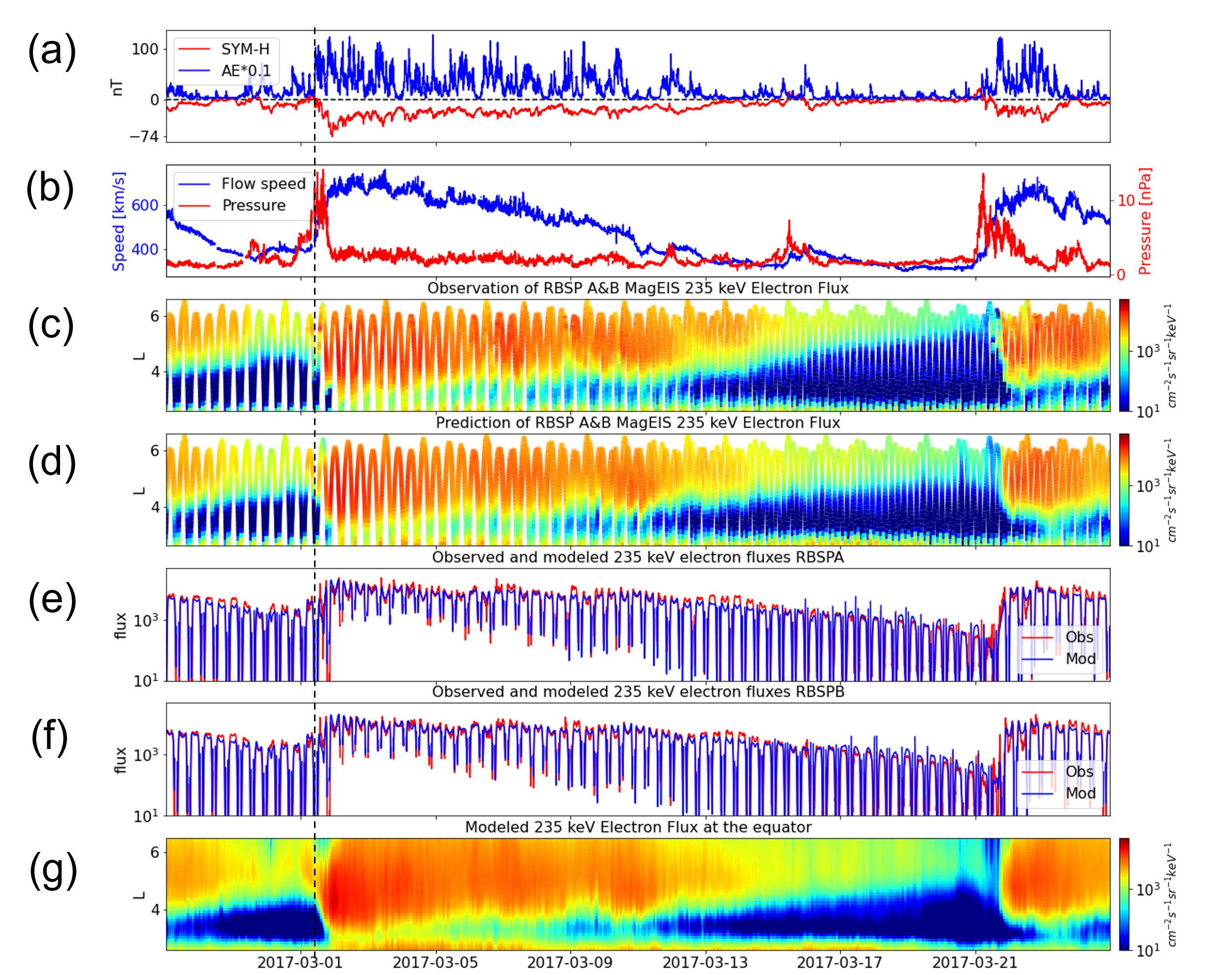}
      \caption{ An example of the 235 keV model results during the month-long period between February 25, 2017, and March 25, 2017, which
      was held out from the training set for test purposes. (a) geomagnetic indices SYM-H and AL; (b) The solar wind flow speed ($V_{sw}$)
      and dynamic pressure ($P_{sw}$); (c-d) the observed and modeled 235 keV electron fluxes as a
      function of L shell and time; (e-f) the observed and modeled 235 keV electron fluxes along the trajectories of
      Van Allen Probe A and B; (g) the modeled 235 keV electron fluxes on the equatorial plane.}
      \label{Figure3}
      \end{figure}

\begin{figure}
      \noindent\includegraphics[width=\textwidth]{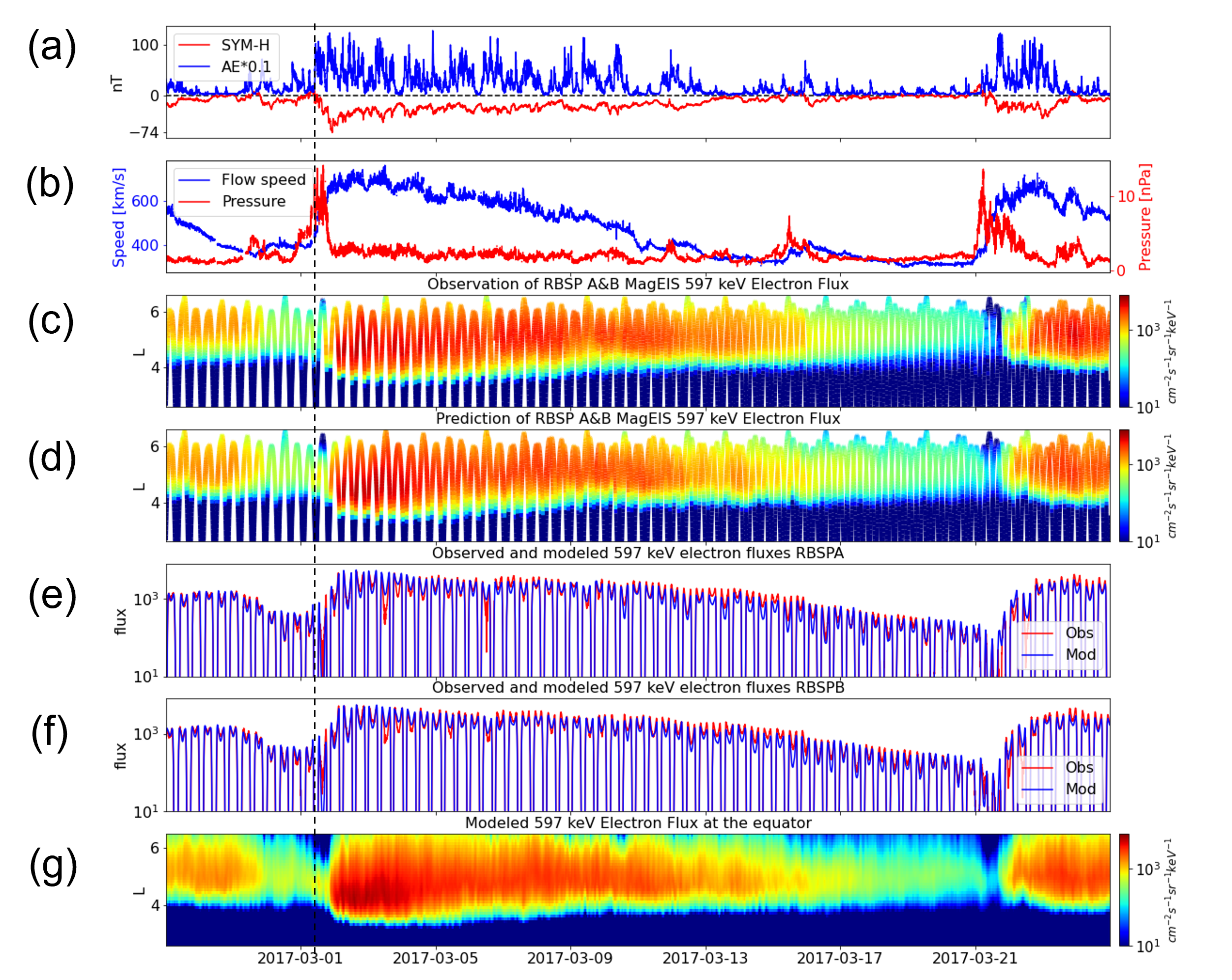}
      \caption{Similar to Figure 3, but for the 597 keV electron fluxes.}
      \label{Figure4}
      \end{figure}
      
A popular metric widely used in machine learning is the $R^2$ score. This indicator is used to quantify the fraction of variance explained by the model. It provides an indication of goodness of fit and therefore a measure of how well unseen samples are likely to be predicted by the model, through the proportion of explained variance. If $\hat{y}$ is the predicted value of the $i$th sample and 
 $y_i$ is the corresponding true value for total $n$ samples, the estimated $R^2$ is defined as:
$$R^{2}(y, \hat{y})=1-\frac{\sum_{i=1}^{n}\left(y_{i}-\hat{y}_{i}\right)^{2}}{\sum_{i=1}^{n}\left(y_{i}-\bar{y}\right)^{2}}$$
where $\bar{y}=\frac{1}{n} \sum_{i=1}^{n} y_{i}$ and this $R^2$ score is the same with prediction efficiency, PE, described in \citeA{reeves2012PE}.

Figure 3 and 4 show a comparison between the observations and model results for two energy channels (235 and 597 keV)
for the out-of-sample one month period between February 25,2017, and March 25, 2017, during which a
moderate geomagnetic storm (Dst minimum -74 nT) occurred. The geomagnetic indices (SYM-H and AE), 
the solar wind dynamic pressure ($P_{sw}$), and flow speed ($V_{sw}$) are shown in panels a and b.
A strong interplanetary shock arrived at ~9 UT on March 1, 2017, as indicated by the substantial
and sudden increase (denoted by the vertical dashed line) in solar wind velocity and solar wind dynamic
pressure.

Figure 3c and 3d show a comparison between the observed and modeled 235 keV electron fluxes as functions 
of time and L shell along the satellite trajectory, showing very close agreement at almost all times.
Figure 3g shows the modeled 235 keV electron fluxes on the equatorial plane (MLAT = $0^{\circ}$, MLT = 0),
highlighting the ANN model's utility in reconstructing electron flux data based on geomagnetic indices
and solar wind parameters alone. 

Figure 4c,4d and 4g show the same analysis but for 597 keV electron
fluxes. Again, the models are seen to accurately reproduce the general variation of the electron fluxes
as well as several key aspects of the general radiation belt behavior, including the effects of physical 
processes such as rapid local acceleration, radial diffusion, and flux decay. In addition, the comparison
of Figure 3 and Figure 4 clearly shows the capabilities of our model in reconstructing the energy-dependent
dynamics of the outer radiation belt.

The analysis in Figure 3 and 4 shows that each model is able to capture the rapid electron flux enhancement
at 2017-03-01 UT and shows excellent agreement in the peak value of the flux and its location.
The model-reproduced enhancement first occurs in the 235 keV channel and is later visible in the 597 keV
channel, which demonstrates that the low energy seed population is gradually accelerated, most likely due
to wave-particle interactions as has been suggested previously \cite{horne2003,jaynes2015,wenli2014}, consistent with observations.
In addition, at 235 keV, the electron fluxes in this enhancement event are seen to fill the 
slot region and penetrate into the inner zone, showing that the enhancement of electrons of the
inner zone is also well captured. At 597 keV and 909 keV (Supplementary Figure 1), the electron fluxes
do not penetrate into the lower L shells, and the ANN model reproduces the "impenetrable barrier"
precisely, implicitly "baking in" all the relevant physical processes into the model that prevent further
penetration of these electrons to lower L-shells \cite{Baker2014,Reeves2016}. Last, the slow decay of the electron
fluxes is well captured. Interestingly, the models successfully reproduce the timing, L-shell, and energy dependence of this
decay process. From March 14, the flux decayed more rapidly at $4.0 < L < 4.5$ for the 235 keV
channel electrons  than it did for 597 keV electrons consistent with previous observations and theoretical
estimates of electron lifetimes. Similar features are also observed before March 1 \cite{Seth2020b,Seth2020a,Reeves2016}.

\begin{figure}
      \noindent\includegraphics[width=\textwidth]{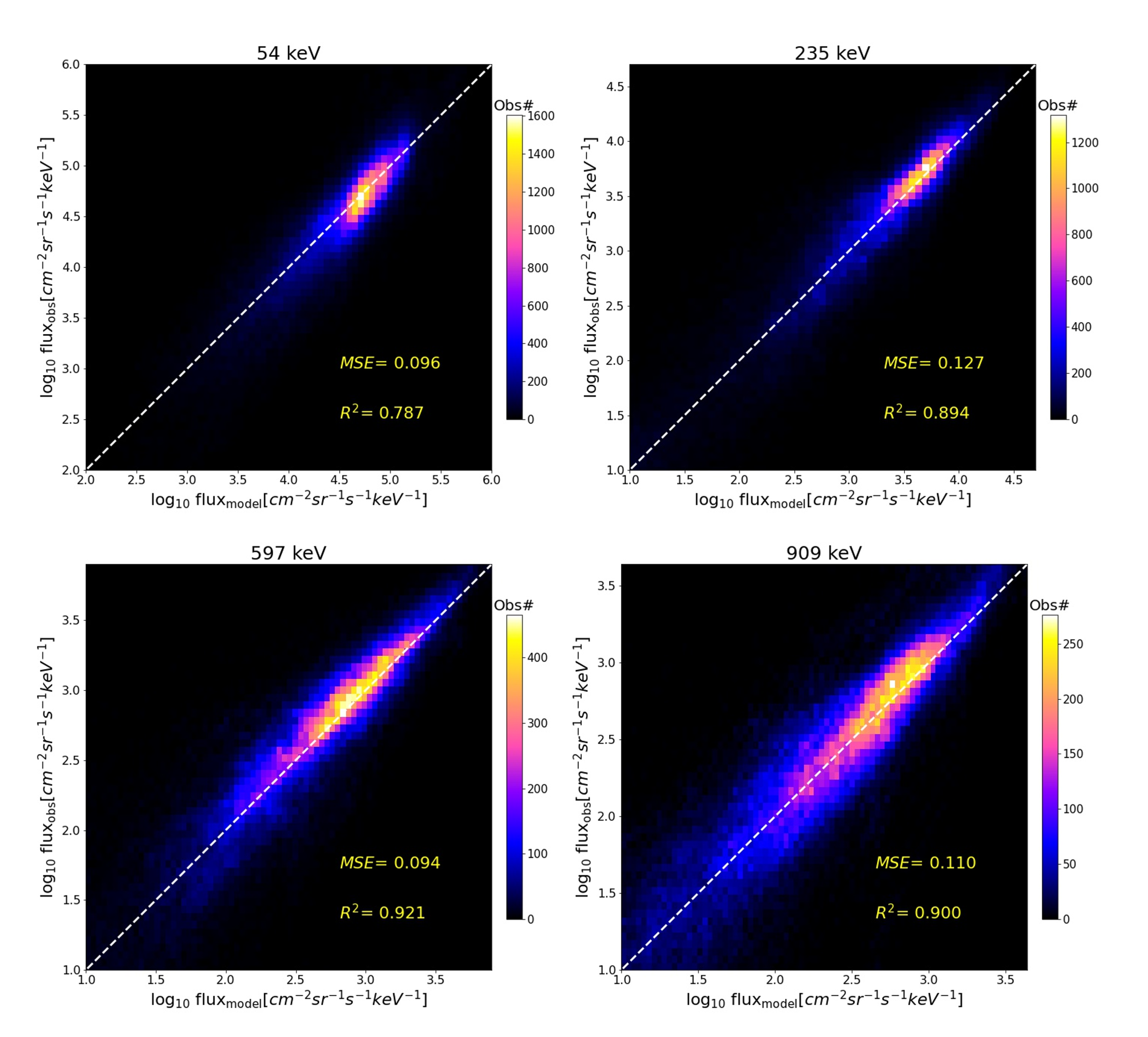}
      \caption{Model performance on the test dataset of Part 3 for different electron energy channels, (a-d) test data at
      54, 235, 597, and 909 keV. The white dashed lines are diagonal lines that indicate perfect agreement (y=x) between
      the observations and model results. The coefficient of determination R-squared and MSE are shown at the right bottom corners.}
      \label{Figure5}
      \end{figure}
Figure 3e and 3f show the comparison between the observed (red) and modeled electron fluxes along
the RBSP-A and RBSP-B trajectories, and a similar analysis is shown for the 597 keV electron fluxes
in the Figure 4e and 4f. Based on this comparison, the {overall }R-squared value of those two channels for the 
out-of-sample period are calculated to be 0.91 (235 keV) and {0.96} (597 keV) and the mean-squared error (MSE) between model results
and observations are 0.10 (235 keV) and {0.05} (597 keV). The out-of-sample performance metrics are consistent with
the performance {of Part 3 test dataset }analysis shown in Figure 5. These results indicate that the models can predict
the (sequentially organized) out-of-sample data with a $90\%$ cross-correlation (i.e., variability of the response data around its mean)
and a factor of $\sim 2$ performance in those channels which are comparable to the instrument accuracy. The performance of the 
54 keV electron fluxes is not as good as the higher energy channels, as expected due to its complicated and rapid dynamics, as 
well as consisting of uncorrected data, which may add uncertainty to observed flux values. But the result in the 
out-of-sample period and test dataset ($R^2 \sim 0.8$) are still satisfactory and capture the important dynamics (See Supplementary Figure S1).

\begin{figure}
      \noindent\includegraphics[width=\textwidth]{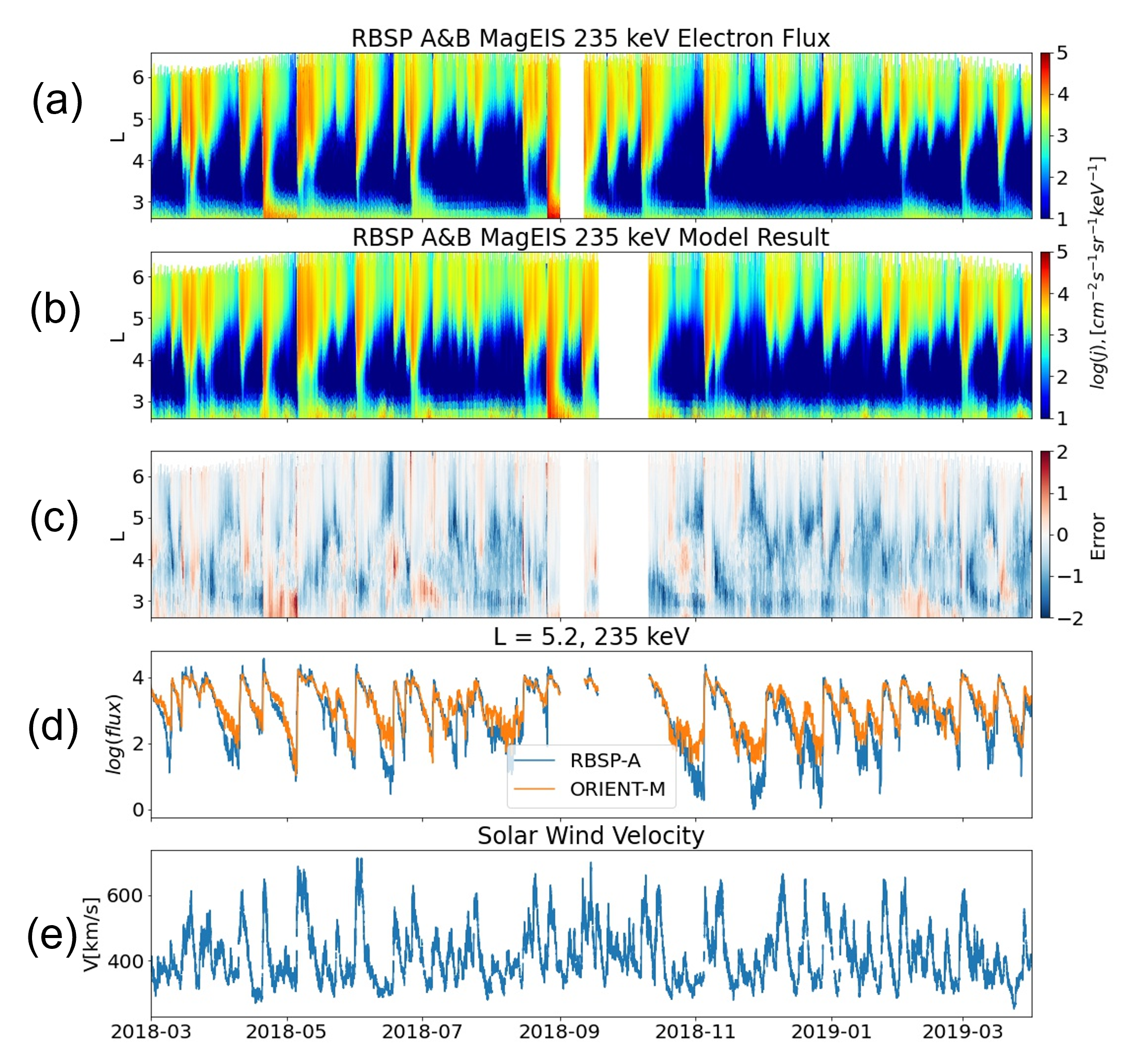}
      \caption{\add{Out-of-sample} model results produced along the Van Allen Probes' trajectories after March 2018 using the predicted AL index (\url{https://lasp.colorado.edu/home/personnel/xinlin.li})
      as input while the other inputs are obtained from the OMNI database. (a) The observed 235 keV 
      electron fluxes along the trajectories of Van Allen Probes. (b) The 235 keV model results along the trajectories.
      (c) The differences between the observed and modeled electron fluxes, which are defined as $log_{10}(flux_{obs} + 1) - log_{10}(flux_{model} + 1)$. (d)  comparison of observed and predicted flux at the fixed L-shell of 5.2 (e) solar wind velocity
      The white gap in panel (a) is due to missing data on the RBSP data website (\url{https://rbsp-ect.lanl.gov/rbsp_ect.php}). The gap in panel
      (b) is due to a 10-day gap in the predicted AL index.}
      \label{Figure6}
      \end{figure}
\begin{figure}
      \noindent\includegraphics[width=\textwidth]{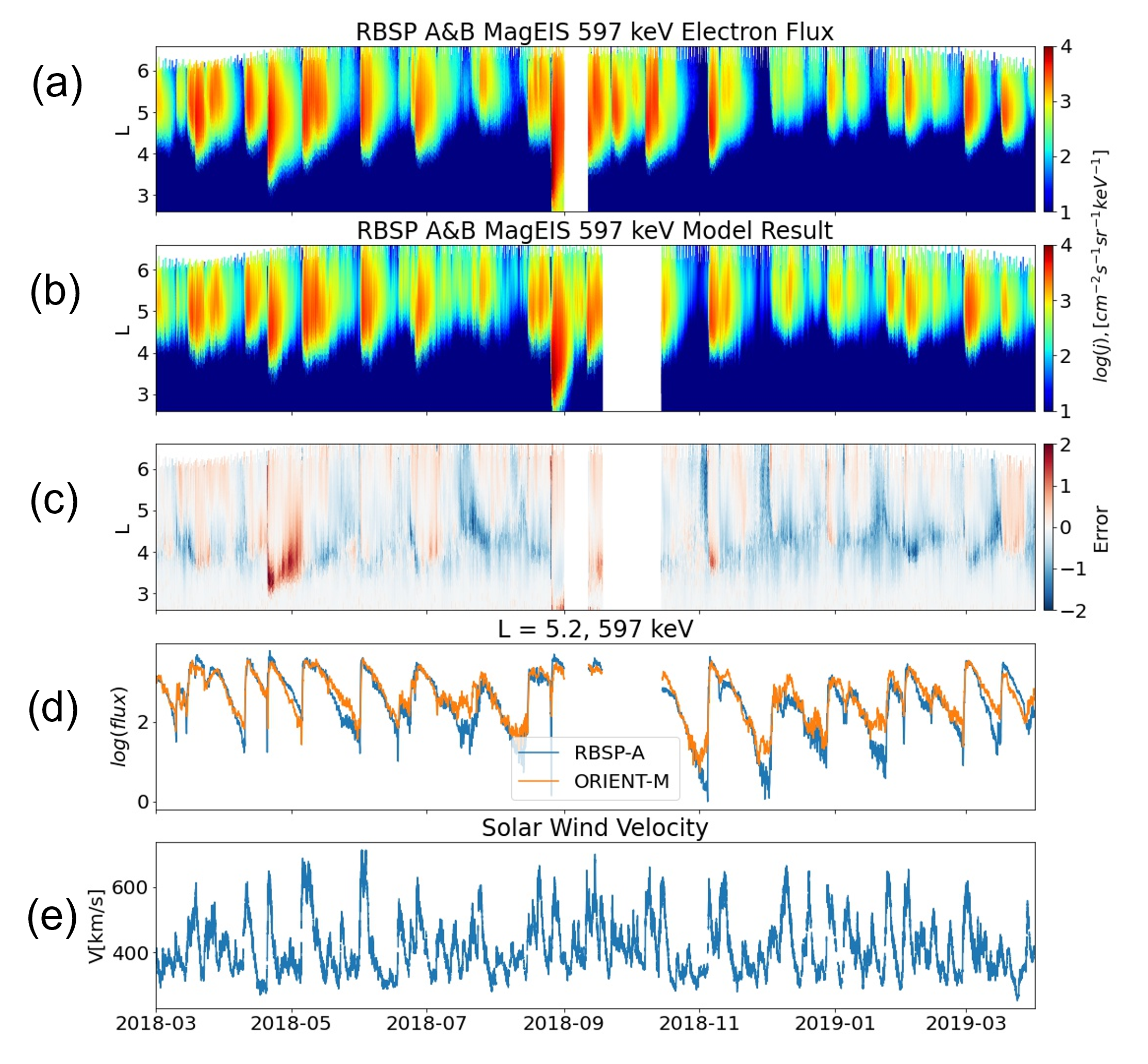}
      \caption{Similar to Figure 6, but for the 597 keV electron fluxes.}
      \label{Figure7}
      \end{figure}
{Figure 6 and 7 (a - c) show comparisons} between the observed and modeled 235 keV and 597 keV electron fluxes 
along the trajectories of RBSP A and B between March 1, 2018 and April 1, 2019. It is important to note that here we are using predicted AL index since the observed OMNI AL index is not available after March 2018. Because the electron
fluxes during this time period were not included in the training process (Section 3.2), the comparisons demonstrate the 
models' out-of-sample predictive ability. The predicted AL index using upstream solar wind measurements is
available at \url{https://lasp.colorado.edu/home/personnel/xinlin.li} \cite{xinlinal,luo}, which has a linear correlation coefficient of 0.846 and a prediction efficiency of 0.715. The errors, as shown in Figure 6c and 7c, are defined as the difference between the training target
of observed and modeled electron fluxes:$log_{10}(flux_{obs}+1) - log_{10}(flux_{model}+1)$. The results show that our models
can reproduce rich dynamics in the fluxes, with rapid enhancements penetrating through the slot region into
the inner zone. In particular, the models capture the rapid enhancement and penetration depth
during the strong storm on Aug 26, 2018. The R-squared value of those two channels for the out-of-sample period {along the orbit}
are calculated to be {0.80} (235 keV) and {0.87} (597 keV) and the mean-squared error (MSE) between model results and observations are {0.2} (235 keV)
and {0.15} (597 keV). We emphasize that the errors originate from both the error in the predicted AL index and from our models.
{Figure 6d and 7d show the RBSP-A observations and the ORIENT-M output for L around 5.2 ( select data from $5.1<L<5.3$). The $R^2$ scores between the two are 0.71 (235 keV) and 0.77 (597 keV). One can therefore conclude that the ORIENT-M model generalizes well on the unseen data, both at a fixed L and along the orbit.} \remove{The $R^2$ values of the four channels' test datasets for different L-shells are shown in the Supplementary Figure S3. It is shown clearly that except the margin region, our models produce good results. The relatively stable inner zone which extends to higher L shell for low energy flux could account for the very low $R^2$ value of the 54 keV channel at $L \sim 3$.}

\add{The $R^2$ values of the four channels' test datasets for different L-shells and MLTs are shown in the Figure 9. It is shown clearly that except the margin regions, our models can produce good results. The 54 keV model performance of the inner zone, which extends to higher L shell for low energy flux, is not as good as the outer zone. A possible explanation may be that the dynamics of inner zone are much less than that of the outer zone and our model cannot capture it accurately. For the regions with low $R^2$ value of 597 keV ($L \sim 3$, MLT from 6 to 8) and 909 keV ($L \sim 3$, MLT from 4 to 8), the samples in the test dataset are only
composed of very low flux measurements ($flux < 10 [cm^{-2}sr^{-1} s ^{-1}keV^{-1}]$) so that these discrepancies can be negligible. }

Since the velocity of the electron drift motion around the Earth is energy-dependent, the global electron distribution following
storms and substorms is expected to show an energy-related variation in MLT. This energy-dependent distribution results from a competition
of the electron drift period and the loss timescale for that particular energy of electrons. For relativistic electrons ($\sim$MeV), the drift period is very small (tens of minutes) and the loss timescale is large (several days),
so the drift motion dominates, and the fluxes are almost symmetrical in MLT about the Earth. However, at lower electron energies (tens to a few hundred keV),
the energy-dependent magnetic drifts are far slower, the drift periods are longer, and the loss timescales are shorter, such that
the electron fluxes can vary dramatically within a limited range of MLTs.
\begin{figure}
      \noindent\includegraphics[width=0.95\textwidth]{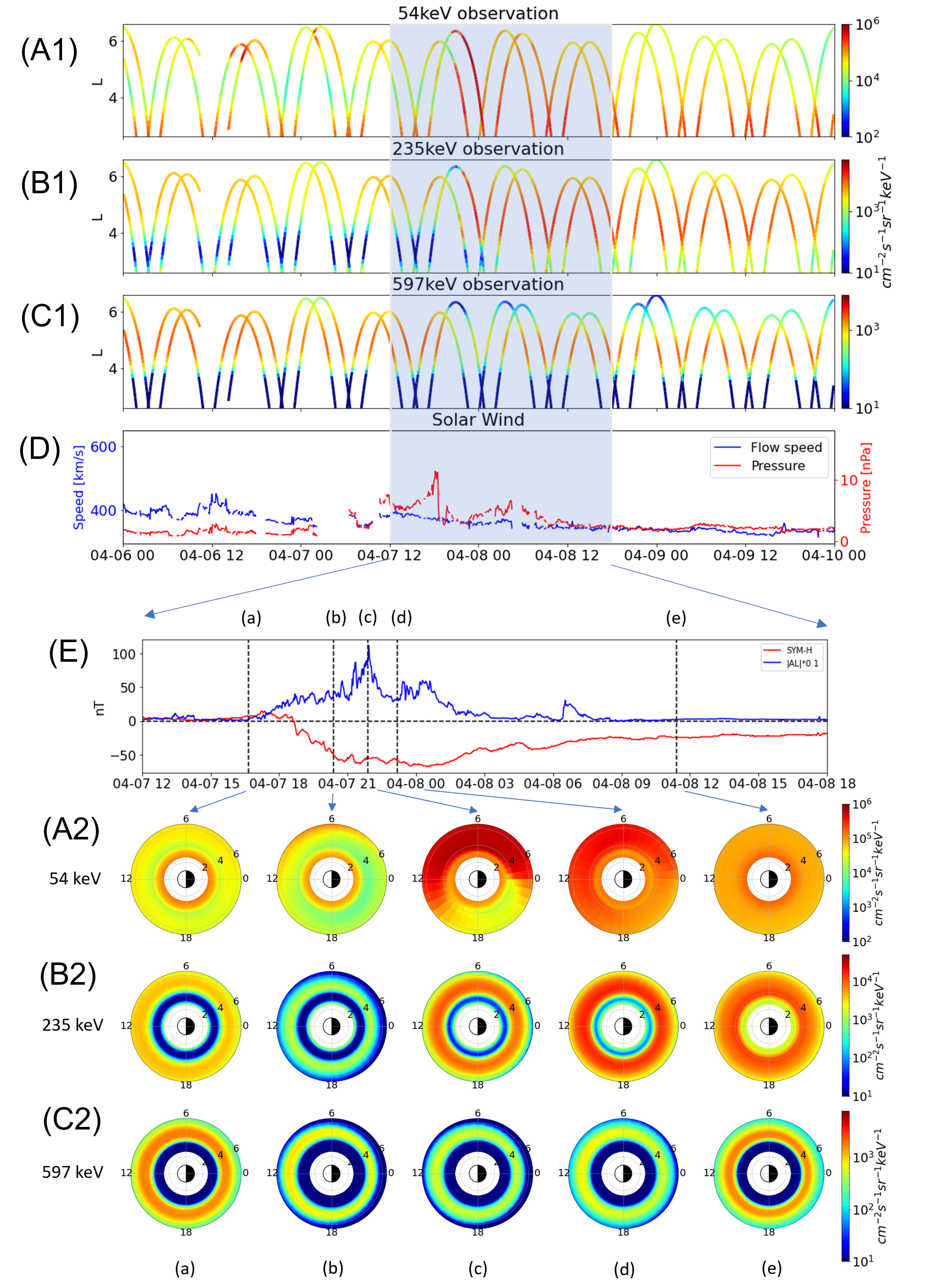}
      \caption{(A1-C1) Observations from 2016-04-06 to 2016-04-10 for 54keV, 235keV and 597keV channel. (D) Solar wind speed and pressure from OMNI. (E) Selected range of SYM-H and AL (2016-04-07/12:00 - 2016-04-08/18:00 ).
      (A2-C2) A series of panels showing the model reconstruction of the electron fluxes as a function
      of L-shell and MLT for different energy channels (54 keV, 235 keV, and 597 keV) at different snapshots in time:
      (a) 2016-04-07/16:40, in the quiet period before the storm, (b) 2016-04-07/20:25, during the main phase,
      (c) 2016-04-07/21:55 the time of maximum $|AL|$, (d) 2016-04-07/23:10, early recovery, and (e) 2016-04-08/11:25, 
      the late recovery period after the storm.The top panel shows the SYM-H and AL index during the course of the storm.}
      \label{Figure8}
      \end{figure}
Figure 8 shows the modeled equatorial electron flux distribution variation for 54 keV, 235 keV, and 597 keV {(A2, B2, C2) and the corresponding RBSP observations (A1, B1, C1)}
during a geomagnetic storm that occurred on April 8, 2016. The {ORIENT-M} model results show the distribution of 
L-shell and MLT, which exhibit different dynamics for each energy channel. Before the storm (time a), the electron fluxes at
each energy channel are seen to be approximately symmetric in MLT. However, when the storm begins (time b), there
is an observed enhancement for the 54 keV electrons at the Earth's dawnside consistent with plasmasheet electrons being convected from the tail and 
drifting Eastwards due to magnetic gradients. In contrast, the higher energy channels exhibit a dropout in fluxes that
has a roughly MLT-symmetric distribution. At the peak of the geomagnetic activity (as measured by a minimum in the AL index) the 54 keV electron fluxes (time c)
show significant MLT asymmetry in the electron flux enhancements, which is observed predominantly at $L \sim 3-6$, and 
is a reflection of the eastward drift of electrons around the Earth, as well as the rapid loss processes (e.g., scattering by chorus waves) that remove the majority
of the enhanced electron fluxes as they drift from the day through to the dusk sectors (See Figure 8 54 keV (d) and supplementary video). \add{It is worth mentioning that the result of Figure 8 cannot be verified at every MLT since we only have two satellites that cannot be everywhere present simultaneously.}

\remove{The $R^2$ score of the model results and observations in Figure 8 cannot be calculated since we don't have in-situ observations at every MLT.} \change{But these}{The} results of localization to the dawnside of the Earth are consistent with a previous study of the 
same observation, a case study, and statistical distribution result \cite{allision2017,zhaohong2017}. 
In the early recovery phase of the storm (time d), the enhanced fluxes begin to diminish, and the distribution begins
to take on a more MLT-symmetric structure. In the late recovery (time e), the electron distributions in all channels 
begin to relax back to their pre-storm values. {The dropout process might result from the magnetopause shadowing as the dynamic pressure increases. }It is interesting to note that the peak flux levels occur at increasingly later times,
with increasing electron energies, which is consistent with previous observations \cite{Thorne2013} and is likely related to the timescales of 
the local acceleration process. It is worth mentioning that the detailed physics of radiation belt dropouts is still an open problem which we do not aim to address in the present study. We can only speculate that since there was a dynamic pressure enhancement when the drop out happened, it is likely that magnetopause shadowing was the main loss process, but interpretation of this physics is not our focus.

Thus, Figure 8 \add{and 9} show that {the ORIENT-M} model is able to reconstruct the dynamic evolution of electron fluxes at different energies
and capture the L-shell and MLT dependence of both the lower energy fluxes as well as the higher energy fluxes, which is challenging
to study using in-situ observations alone.

\begin{figure}
      \noindent\includegraphics[width=\textwidth]{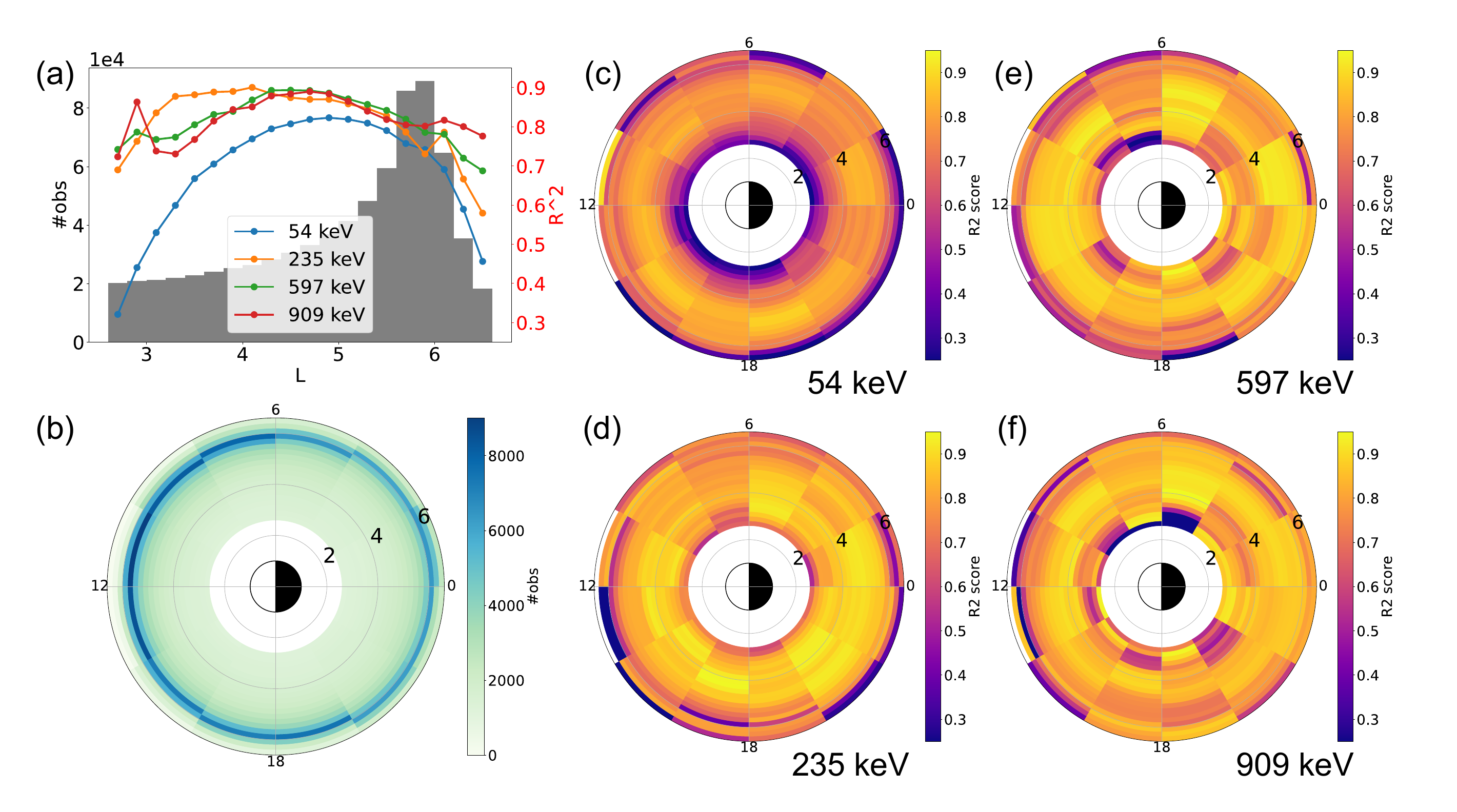}
      \caption{R2 score in the test dataset. (a): R2 values of test datasets for different L shells. The background gray hist data
    shows all the data distribution along L shell of 54 keV channel. (b) The training data distribution of 54 keV. (c) -(f): The R2 values of test dataset as function of L and MLT for different channels. The result includes 20 Lshell bins from $L = 2.6$ to $L = 6.6$ with $\Delta L = 0.2$ and 12 MLT bins with $\Delta MLT = 2$.}
      \label{Figure9}
      \end{figure}

\section{Summary}
Here, we presented a set of Outer Radiation belt Electron Neural net model for Medium energy electrons (ORIENT-M).
The ORIENT-M models are driven purely by a time history of geomagnetic indices and solar wind parameters, meaning that they do not need
any other in situ data (e.g., fluxes from LEO satellites) to provide boundary conditions as inputs into the model, and thus our models are 
able to reconstruct energetic electron fluxes for periods deep in the past and into the future, as long as the geomagnetic indices and solar wind parameters
are available. Our models are trained using six years of electron flux data from the MagEIS instrument onboard
Van Allen Probes, and we have shown only four representative energy channels in this study. Any predictions of our models for periods in the 
past or the future are thus essentially a prediction of what the MagEIS instrument on Van Allen Probes should observe under the input conditions
provided to the model at that time.

The model performance of the test dataset (which was held out during model training), as well as for complete out-of-sample periods, i.e., a geomagnetic storm in March 2017,
was demonstrated and shows a high coefficient of determination {($R^2 \sim 0.7 - 0.9$)}. The model could reproduce the effects of particle acceleration,
decay of the electron fluxes, and energy-dependent dynamics of electron fluxes in the radiation belt. Given its excellent performance in predicting low to high
energy electron fluxes, our model was able to capture the MLT-dependent electron flux dynamics. We presented the MLT 
dependence during a geomagnetic storm that occurred on April 8, 2016. The timing and MLT differences of electron enhancement at 54 keV
were well captured by the model, and hence the implication is that the model baked in all the relevant physical processes that shaped the 
resulting electron distribution. The electron flux asymmetry was seen to be much higher in the dawn sector and the eastward diffusion drift process was reflected in the 
model result.

The ORIENT-M model presented in this study represents a novel capability in modeling the subrelativistic (10s to 100s keV) electron flux
population. At present, such electron fluxes are studied with either first-principles physics-based numerical modeling \cite{Jordanova2010,wenli2010},
neither of which is able to capture the energy-dependent, dynamical variation of the global distribution of the observations.
The present model is aimed at filling this important gap in our modeling capabilities and can be used as an important tool for both space weather
(i.e., spacecraft charging) applications, as well as to discern the physical processes that control the dynamics of this key particle population.

\acknowledgments
The authors would like to thank the NASA SWO2R award 80NSSC19K0239 
for their generous support for this project {as well as subgrant 1559841 to the University of California, Los Angeles, from the University of Colorado Boulder under NASA Prime Grant agreement 80NSSC20K1580}. XC would like to thank grant 
NASA ECIP 80NSSC19K0911 and NASA LWS 80NSSC20K0196. JB acknowledges support 
from the Defense Advanced Research Projects Agency under Department of the Interior 
award D19AC00009. We gratefully acknowledge the MagEIS team, Van Allen Probe mission 
(\url{rbspgway.jhuapl.edu}), predicted AL from Xinlin Li (\url{https://lasp.colorado.edu/home/personnel/xinlin.li}) and OMNI 
dataset (\url{omniweb.gsfc.nasa.gov}). Work at The Aerospace Corporation was supported by 
RBSP-ECT funding provided by JHU/APL contract 967399 under National Aeronautics and 
Space Administration (NASA)'s Prime contract NAS501072. The data and model files are available at \url{https://doi.org/10.5281/zenodo.6299967} and example code is available at \url{https://github.com/donglai96/ORIENT-M}


%
%

\bibliography{ agusample }

%
%
%
%
%

\end{document}


%
%


\title{Supporting Information for "Modeling the dynamic variability of sub-relativistic outer radiation belt electron fluxes using machine learning"}
%
%

%
%



\authors{Donglai Ma\affil{1},Xiangning Chu\affil{2},Jacob Bortnik\affil{1},Seth G.Claudepierre\affil{1,4}
,W.Kent Tobiska\affil{3},Alfredo Cruz\affil{3},S.Dave Bouwer\affil{3},J.F.Fennell\affil{4},J.B.Blake\affil{4}}

\affiliation{1}{Department of Atmospheric and Oceanic Sciences, University of California, Los Angeles}
\affiliation{2}{Laboratory for Atmospheric and Space Physics, University of Colorado Boulder, Boulder, CO, USA}

\affiliation{3}{Space Environment Technologies, Pacific Palisades, CA, USA
}
\affiliation{4}{Space Sciences Department, The Aerospace Corporation, El Segundo, CA, USA
}


%
%

%

\begin{article}

%
%

\noindent\textbf{Contents of this file}
\begin{enumerate}
\item Table of Hyperparameter optimization.
\item Figures S1 and S2

\end{enumerate}
\noindent\textbf{Additional Supporting Information (Files uploaded separately)}
\begin{enumerate}

\item Captions for Movies S1

\end{enumerate}

\noindent\textbf{Introduction}
This supporting information include the description and table of the hyperparameter optimization, two figures, and an additional movie. The figures show the data model comparison for the 54 keV and 909 keV similar to Figures 3 and 4 in the paper. The movie shows the modeled equatorial distributions of the 54 keV flux during a storm. 


\noindent\textbf{Hyperparameter optimization}
The table below shows the search range and the optimal results of the models' hyperparameters for each energy. The hyperparameter optimization is performed using Optuna, which is an open source framework to automate hyperparameter search(\url{https://optuna.org/}). The model is trained using tensorflow(\url{https://www.tensorflow.org/}) which is an open source machine learning platform.

\hfill \break
{\begin{tabular}{|c|c|c|c|c|c|}

Model structure  & Optuna\_range & 54 keV & 235 keV & 597 keV & 909 keV \\ \hline
num\_layers      & (3,4)         & 4      & 4       & 4       & 4       \\ \hline
n\_units\_layer0 & (128,512)     & 291    & 412     & 432     & 428     \\ \hline
dropout\_rates0   & (0.01,0.4)    & 0.211  & 0.368   & 0.011   & 0.382   \\ \hline
n\_units\_layer1 & (4,128)       & 119    & 105     & 114     & 95      \\ \hline
dropout\_rates1  & (0.01,0.4)    & 0.385  & 0.091   & 0.092   & 0.075   \\ \hline
n\_units\_layer2 & (4,128)       & 4      & 116     & 93      & 95      \\ \hline
n\_units\_layer3 & (4,128)       & 29     & 128     & 12      & 120     \\ \hline
\end{tabular}
}

\hfill \break
\noindent\textbf{Table S1.} The Optuna hyperparameter search range and final results for each energy. The 'num layers' represents the number of hidden layers in the fully-connected neural network. The 'n units' stands for the neuron numbers for each layer. The 'dropout ratesx' is the dropout rate for the corresponding layer. The ranges in brackets mean a float number is suggested from min to max for each trial. The final hyperparameters are selected from 200 trials based on results on validation dataset for each channel.

%




\noindent\textbf{Movie S1.} 
The ANN  modeled equatorial electron flux distribution variation for  54 keV channel during a geomagnetic storm that occurred on April 8, 2016.
\end{article}
\clearpage
\begin{figure}

\noindent\includegraphics[width=\textwidth]{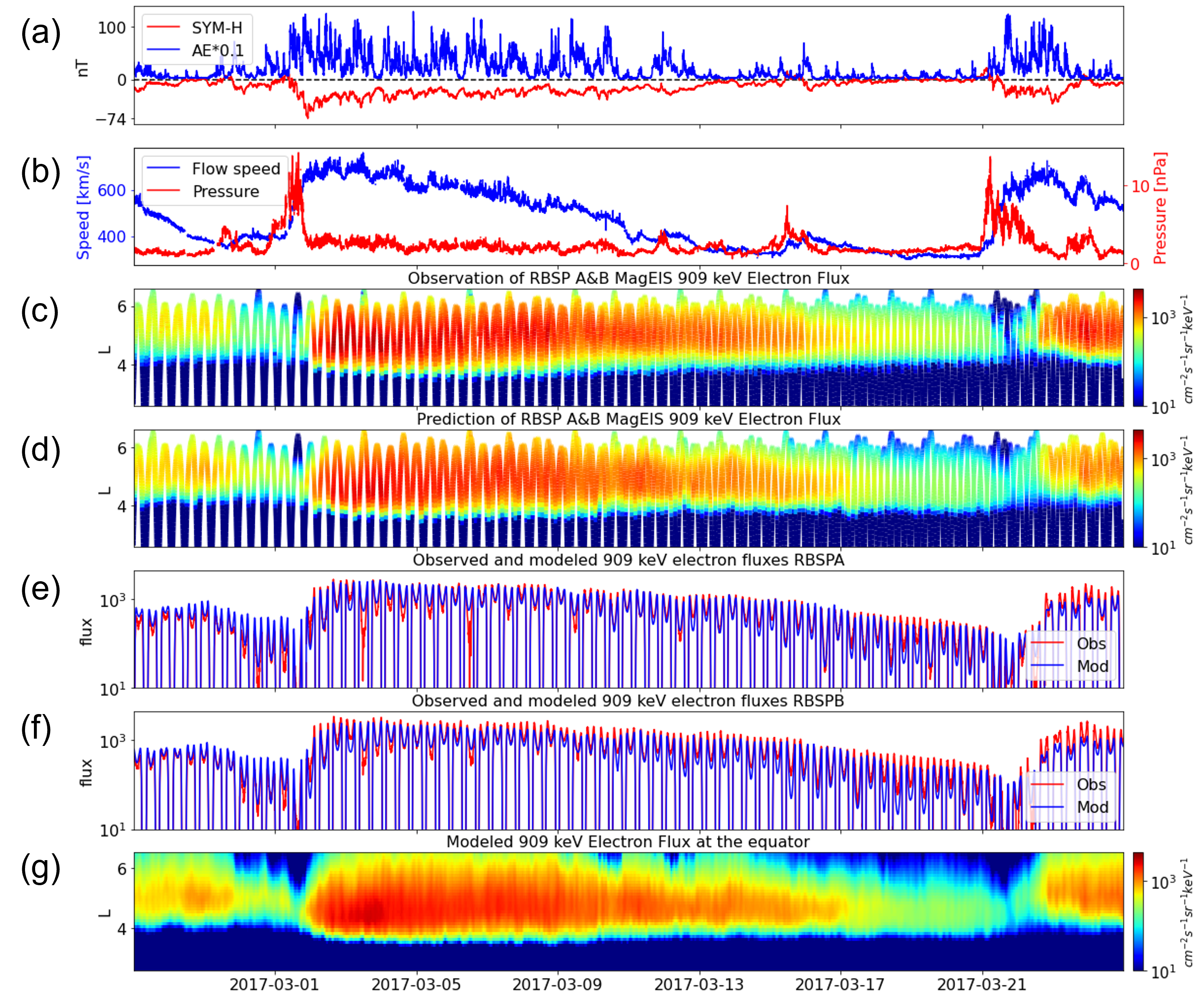}
\caption{An example of the 909 keV model results during the month-long period between February 25, 2017, and March 25, 2017, which was held out from the training set for test purposes. (a) geomagnetic indices SYM-H and AL; (b) The solar wind flow speed ($V_{sw}$) and dynamic pressure ($P_{sw}$);(c-d) the observed and modeled 909 keV electron fluxes as a function of L shell and time; (e-f) the observed and modeled 909 keV electron fluxes along the trajectories of Van Allen Probe A and B; (g) the modeled 909 keV electron fluxes on the equatorial plane.}

\end{figure}
\begin{figure}
\noindent\includegraphics[width=\textwidth]{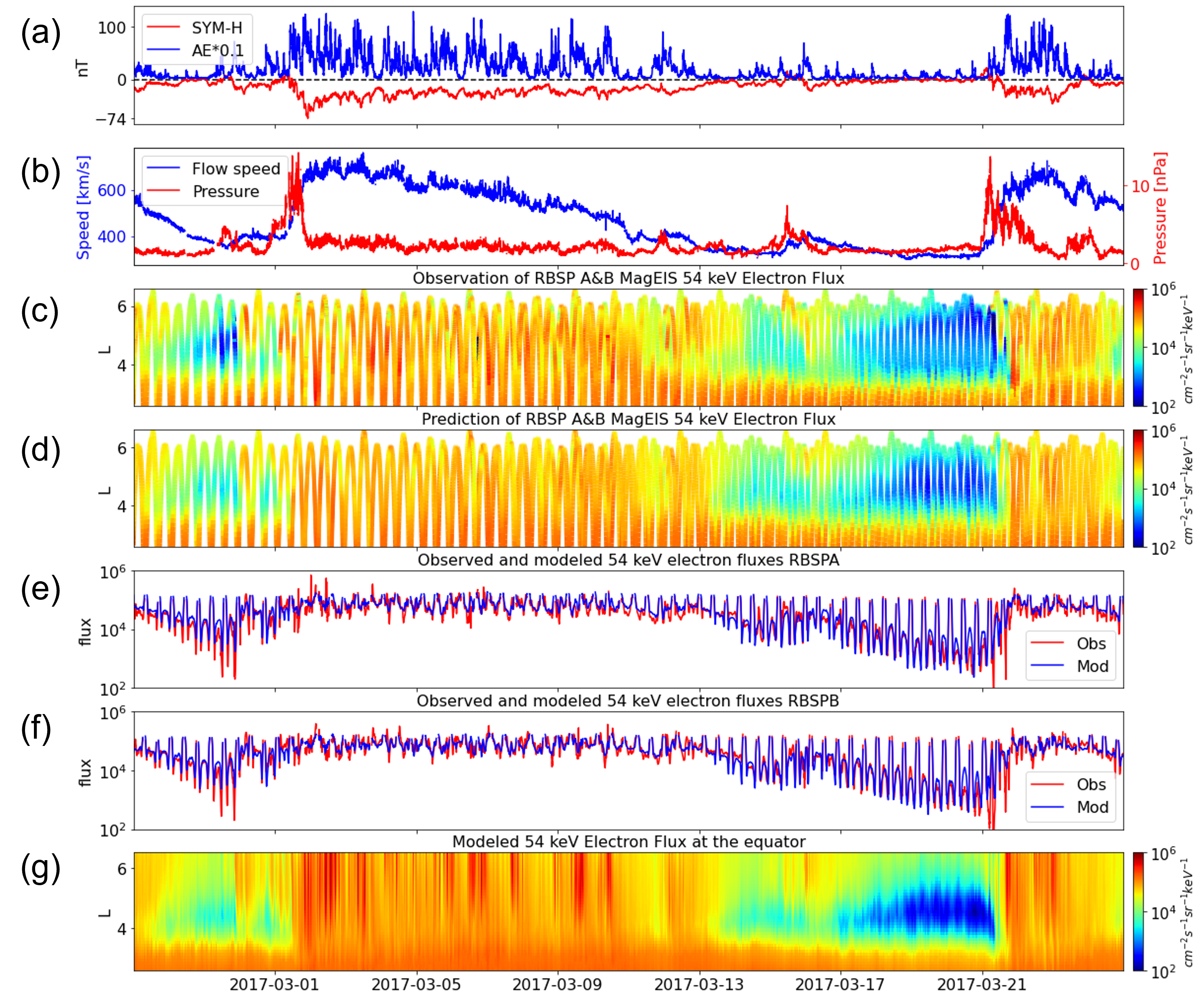}
\caption{Similar to Figure S1, but for the 54 keV electron fluxes.}

\end{figure}